\documentclass[letterpaper,11pt]{article}%\pdfoutput=1 % if your are submitting a pdflatex (i.e. if you have
\usepackage{jheppub} % for details on the use of the package, please
\DeclareFontEncoding{U}{}{}
\DeclareFontFamily{U}{bbold}{}
\DeclareFontShape{U}{bbold}{m}{n}
 {  <5> <6> <7> <8> <9> gen * bbold
   <10> <10.95> bbold10
  <12> <14.4> bbold12
 <17.28> <20.74> <24.88> bbold17
  }{}
\DeclareSymbolFont{bbold}{U}{bbold}{m}{n}
\DeclareSymbolFontAlphabet{\mathbb}{bbold}

\newcommand{\dd}{\text{d}}

\usepackage[export]{adjustbox} % need that for the \makebox command
\newcommand\GravitonPropagator{ \makebox{\raisebox{-0.2cm}{\includegraphics{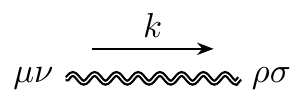}~}} }

\newcommand\GravitonExchange{ \makebox{\raisebox{-1.4cm}{\includegraphics{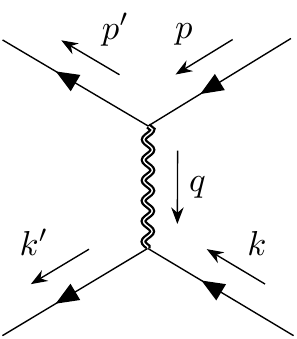}~}} }
\newcommand\OneGravitonOneFermionVertex{ \makebox{\raisebox{-0.75cm}{\includegraphics{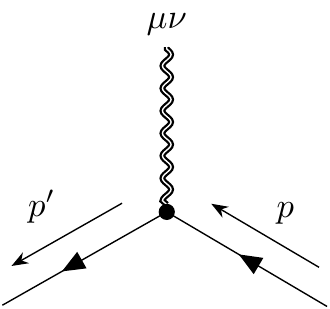}~}} }

\newcommand\OneGravitonOneScalarVertex{ \makebox{\raisebox{-0.75cm}{\includegraphics{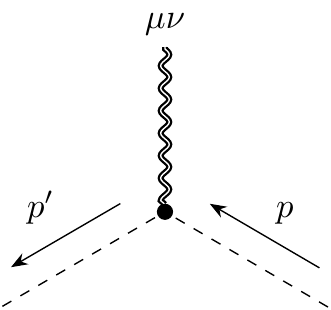}~}} }
\newcommand\TwoGravitonTwoScalarsVertex{ \makebox{\raisebox{-1.55cm}{\includegraphics{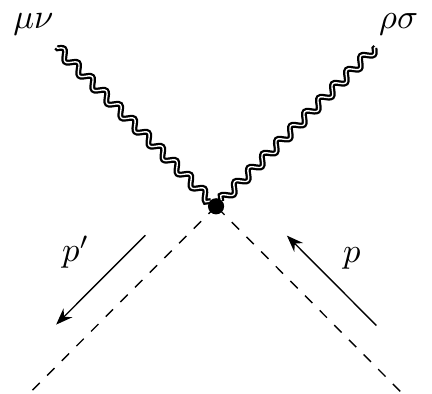}~}} }

\newcommand\BubbleDiagram{ \makebox{\raisebox{-0.75cm}{\includegraphics{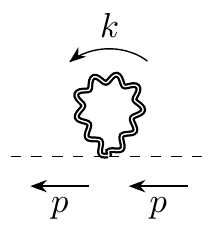}~}} }
\newcommand\RainbowDiagram{ \makebox{\raisebox{-0.85cm}{\includegraphics{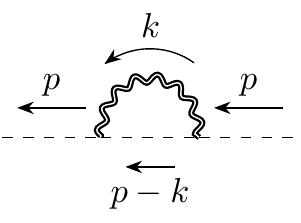}~}} }

\title{Asymptotically nonlocal gravity}

\author[a]{Jens Boos}
\author[a]{and Christopher D. Carone}

\affiliation[a]{High Energy Theory Group, Department of Physics,\\
William \& Mary, Williamsburg, VA 23187-8795, USA}

\emailAdd{jboos@wm.edu}
\emailAdd{cdcaro@wm.edu}

\abstract{Asymptotically nonlocal field theories interpolate between Lee-Wick theories with multiple propagator poles, and ghost-free nonlocal theories.  Previous work
on asymptotically nonlocal scalar, Abelian, and non-Abelian gauge theories has demonstrated the existence of an emergent regulator scale that
is hierarchically smaller than the lightest Lee-Wick partner, in a limit where the Lee-Wick spectrum becomes dense and decoupled.  We generalize this 
construction to linearized gravity, and demonstrate the emergent regulator scale in three examples: by studying the resolution of the singularity (i) at the origin in
the classical solution for the metric of a point particle, and (ii) in the nonrelativistic gravitational potential computed via a one-graviton exchange amplitude; (iii) we also
show how this derived scale regulates the one-loop graviton contribution to the self energy of a real scalar field. We comment briefly on the generalization 
of our approach to the full, nonlinear theory of gravity.}
\keywords{Models of Quantum Gravity, Spacetime Singularities, Effective Field Theories}
\begin{document} 

\maketitle
\flushbottom

\section{Introduction}\label{sec:intro} 

Higher-derivative quantum field theories have been of considerable interest as possible solutions to the hierarchy problem in the 
Standard Model and as a strategy for constructing perturbatively renormalizable quantum theories of gravity.
In this paper, we focus on Lee-Wick theories~\cite{LeeWick:1969,LeeWick:1972,Cutkosky:1969fq}; explicit constructions
addressing the hierarchy problem appear in Refs.~\cite{Grinstein:2007mp,Carone:2008iw} and phenomenological issues have been
studied in Ref.~\cite{LWpheno}.   Higher-derivative quadratic terms 
lead to propagators that fall off more quickly with momentum, so that loop amplitudes typically become more convergent.  
As a result, it has been shown  in  Lee-Wick extensions of the Standard Model that the otherwise quadratic divergence of the Higgs 
boson squared mass become logarithmic, eliminating the fine-tuning needed to keep the Higgs boson light~\cite{Grinstein:2007mp,Carone:2008iw}.  

The higher-derivative quadratic terms of Lee-Wick theories necessarily imply the existence of additional propagator poles, corresponding to heavy partner particles.  In 
the minimal Lee-Wick Standard Model~\cite{Grinstein:2007mp}, these additional poles have wrong-sign residues, corresponding to states with negative norms.  Nonetheless, 
approaches to quantizing these theories consistently while preserving unitarity exist; for the original proposals and their application, see 
Refs.~\cite{LeeWick:1969,LeeWick:1972,Cutkosky:1969fq,Grinstein:2008bg}, and for more modern treatments that address their ambiguities see 
Refs.~\cite{Anselmi:2017yux,Anselmi:2017lia,Anselmi:2018kgz,Chin:2018puw}.   Like other models that have partner particles that are of interest in addressing the hierarchy problem (for example, 
the Minimal Supersymmetric Standard Model), fine-tuning is reintroduced if the partners are taken to be heavy.   The non-observation of partner particles at the Large Hadron Collider, 
which now probes scales much heavier than the Higgs boson mass, motivates us to consider whether one can construct theories where the Lee-Wick partners can be sufficiently 
decoupled from the low-energy effective theory without reintroducing a fine-tuning problem in a light scalar mass.

Asymptotically nonlocal theories, which are the focus of the present work, are Lee-Wick theories that have this desired property~\cite{Boos:2021chb,Boos:2021jih,Boos:2021lsj}.  These theories 
interpolate between higher-derivative theories of finite order and nonlocal theories that are ghost-free~\cite{Efimov:1967,Krasnikov:1987,Kuzmin:1989,Tomboulis:1997gg,Modesto:2011kw,Biswas:2011ar,Ghoshal:2017egr,Buoninfante:2018mre,Boos:2020qgg,Ghoshal:2020lfd} via a sequence of theories with $N-1$ massive partner particles, as $N$ becomes 
large. (Following our past conventions, $N$ refers to the total number of propagator poles~\cite{Boos:2021chb,Boos:2021jih,Boos:2021lsj}.)  In the limiting case 
where $N \rightarrow \infty$ with the ratio $m_1^2/N$ held fixed, where $m_1$ is the mass of the lightest Lee-Wick particle, one arrives at a nonlocal theory whose quadratic terms 
involve the form factor $\exp(\ell^2\Box)$.   In such a theory, the scale $\ell^{-1}$ serves as a regulator.  At large but finite $N$, a derived regulator scale
$\ell^2 \sim {\cal O}( N/m^2_1)$, which does not appear as a fundamental parameter in the Lagrangian, emerges in physical quantities and as the regulator of loop amplitudes.
Since this dervied scale can be hierarchically separated from the mass scale of the lightest Lee-Wick particle, even at finite $N$, the hierarchy problem can be addressed in realistic 
Lee-Wick theories when partner particles are much heavier than the scalar mass that one would like to protect~\cite{Boos:2021jih,Boos:2021lsj}.   Asymptotically nonlocal theories have 
been studied in the context of $\phi^4$-theory~\cite{Boos:2021chb},  scalar quantum electrodynamics~\cite{Boos:2021jih}, and non-Abelian gauge theories~\cite{Boos:2021lsj}.  
If this approach has relevance to addressing the hierarchy problem of the Standard Model, it is natural to ask whether the gravitational sector could have a similar structure.
In the present work, we show how the approach of Refs.~\cite{Boos:2021chb,Boos:2021jih,Boos:2021lsj} may be extended to gravity, working perturbatively at lowest order in an expansion of 
the metric about a flat background.  Lee-Wick generalizations of Einstein gravity are interesting in their own right and have been discussed before in the literature as potential quantum theories of gravity \cite{Tomboulis:1977,Tomboulis:1980,Modesto:2016a,Modesto:2016b}.

Our paper is organized as follows. We first review the concept of asymptotic nonlocality in a simple scalar toy model in Sec.~\ref{sec:framework}, and present a field theory for 
asymptotically nonlocal gravity at the linearized level in Sec.~\ref{sec:gravity}.  In Sec.~\ref{sec:examples}, we demonstrate the emergence of the nonlocal regulator scale in our 
gravitational theory via three examples:  (i) the resolution of the singularity in the classical solution for the metric of a point particle, (ii) the same for the nonrelativistic potential extracted 
from a $t$-channel graviton exchange scattering amplitude, and (iii) the ultraviolet behavior of the gravitational contributions to the self-energy of a real scalar field at one loop.  In 
Sec.~\ref{sec:conc}, we summarize our results and  comment on how to extend these studies beyond the linearized approximation by drawing 
analogies to the construction of asymptotically nonlocal Yang-Mills theory~\cite{Boos:2021lsj}. 

\section{Framework} \label{sec:framework}

In this section, we review the construction of an asymptotically nonlocal theory of a real scalar field.   The goal is to 
find a sequence of higher-derivative quantum field theories, each with a finite number of derivatives, that approach the nonlocal theory defined by
\begin{equation}
{\cal L} = -\frac{1}{2} \, \phi \, \Box \, e^{\ell^2 \Box} \, \phi - V(\phi) 
\label{eq:limitpoint}
\end{equation}
as a limit point.  The existence of such a limit is desirable so that a derived regulator scale similar to $\ell^{-1}$ emerges in the theories with 
large but finite $N$ that are of interest to us. One begins by noting that 
\begin{equation}
{\cal L} = -\frac{1}{2} \, \phi \,\, \Box \, \left(1+\frac{\ell^2 \Box}{N-1}\right)^{N-1} \!\! \phi - V(\phi) 
\label{eq:toodegen}
\end{equation}
approaches Eq.~(\ref{eq:limitpoint}) in the limit that $N\rightarrow \infty$.  However, the propagator following from Eq.~(\ref{eq:toodegen}) 
includes an $(N-1)^{\rm th}$ order pole, which has no simple particle interpretation.  We remedy this by altering the finite-$N$ theory:
\begin{equation}
\label{eq:hdN}
{\cal L}_N = - \frac{1}{2} \phi \, \Box \left[\prod_{j=1}^{N-1} \left(1+\frac{\ell_j^2 \Box}{N-1} \right) \right] \phi - V(\phi) \, .
\end{equation}
 One obtains the same limiting form of the Lagrangian, Eq.~(\ref{eq:limitpoint}), when  $N\rightarrow \infty$, provided that the $\ell_j$ 
 approach a common value, $\ell$, as the limit is taken.  For finite-$N$, the propagator is 
now given by
\begin{equation}
D_F(p^2) = \frac{i}{p^2} \, \prod_{j=1}^{N-1} \left(1-\frac{\ell_j^2 \, 
p^2}{N-1}\right)^{-1} \, ,
\label{eq:nondegprop}
\end{equation}
which has $N$ nondegenerate poles.  Let us define $m_0=0$ and $m_k = 1/a_k$ where $a_k^2 \equiv \ell_k^2 / (N-1)$ for $k \geq 1$.
Then, one may use a partial fraction decomposition to write Eq.~(\ref{eq:nondegprop}) as
\begin{equation}
D_F(p^2) = \sum_{j=0}^{N-1} c_j \, \frac{i}{p^2 -  m_j^2} \, ,
\label{eq:sumprops}
\end{equation}
where
\begin{equation}
c_0=1 \,\,\,\,\, \mbox{ and } \,\,\,\,\, 
c_j =  -\prod\limits_{\substack{k=1\\k\not=j}}^{N-1}  \frac{m_k^2}{m_k^2 - m_j^2}  \, \mbox{ for } j>0  \, .
 \label{eq:pfd0}
\end{equation} 
It follows from Eq.~(\ref{eq:pfd0}) that the residue of each pole alternates in sign, indicating that the spectrum of the theory consists of 
a tower of ordinary particles and ghosts.   This is what one expects to find~\cite{Pais:1950za} in a Lee-Wick theory with 
additional higher-derivative quadratic terms beyond the minimal set~\cite{Carone:2008iw}.  

While  the theories defined by Eq.~(\ref{eq:hdN}) at any finite $N$ are of the Lee-Wick type, they inherit a 
desirable feature of the limiting theory as $N$ becomes large, namely the emergence of a derived nonlocal scale that can serve as a regulator. 
The relationship between the regulator scale and the mass of the lightest Lee-Wick resonance, an observable quantity, is of fundamental interest in 
Lee-Wick theories~\cite{Grinstein:2007mp}.  We showed in Refs.~\cite{Boos:2021chb,Boos:2021jih,Boos:2021lsj}  that the nonlocal scale, 
$M_\text{nl} \equiv 1/\ell$,  can be hierarchically smaller than the lightest Lee-Wick partner $m_1$,   
\begin{align}
M_\text{nl}^2 \sim {\cal O}\left(\frac{m_1^2}{N}\right) \, .
\end{align}
A parametric suppression of the regulator scale has been found in other constructions discussed in the 
literature~\cite{Dvali:2007hz,Buoninfante:2018gce}.   As we will see, one encounters the same phenomenon in the linearized gravitational theory that is the 
subject of the present work.

The existence of $N$ first-order poles in Eq.~(\ref{eq:sumprops}) suggests that there is a way to rewrite Eq.~(\ref{eq:hdN}) as a theory 
that includes $N$ propagating fields that do not have higher-derivative quadratic terms.   This approach is well known in Lee-Wick 
theories~\cite{Grinstein:2007mp,Carone:2008iw}, and was extended to asymptotically nonlocal theories in Refs.~\cite{Boos:2021chb,Boos:2021jih}.   
In the present case,  consider the following theory of $N$ real scalar 
fields $\phi_j$, and $N-1$ real scalar fields $\chi_j$:
\begin{equation}
{\cal L}_N = -\frac{1}{2} \, \phi_1 \Box \phi_N - V(\phi_1) - \sum_{j=1}^{N-1} \chi_j \, \left[ \Box \phi_j - (\phi_{j+1}-\phi_j)/a_j^2\right] \, .
\label{eq:start}
\end{equation}
The $a_j$ were defined previously, and we have rescaled the $\chi_j$ fields so that each term in the sum has unit coefficient.
The $\chi_j$ are auxiliary fields that serve to impose a set of constraints on the theory.  Since they appear linearly in the Lagrangian, one may
functionally integrate over the $\chi_j$ in the generating functional for the correlations functions of the theory.  This leads to functional delta functions, 
which impose constraints that are exact at the quantum level:
\begin{equation}
\Box \phi_j -(\phi_{j+1}-\phi_j)/a_j^2=0 \, , \quad \mbox{ for }j=1, \dots, N-1 \, .
\label{eq:recursive}
\end{equation}
These constraints allow the $(j+1)^{\rm th}$ field to be eliminated in terms of the $j^{\rm th}$ field; after successive functional integrations, 
one finds
\begin{equation}
\phi_N = \left[\prod_{j=1}^{N-1} \left(1+\frac{\ell_j^2 \Box}{N-1} \right) \right] \phi_1 \, .
\end{equation}
Substituting into Eq.~(\ref{eq:start}), and relabelling $\phi_1 \rightarrow \phi$, one recovers Eq.~(\ref{eq:hdN}), as desired.

Alternatively, Eq.~(\ref{eq:start}) can be subject to field redefinitions which lead to a sector with diagonal kinetic and mass terms, corresponding to the expected propagating degrees of freedom in a Lee-Wick theory, and a sector of non-dynamical fields that can be integrated out. The spectrum
of propagating fields is found to be identical to that of the higher-derivative theory; we refer the reader to Ref.~\cite{Boos:2021chb}  for details. In applications where one is only interested in computing Feynman diagrams with internal scalar lines, there is little practical 
advantage to using a field-redefined version of Eq.~(\ref{eq:start}) instead of the higher-derivative form in Eq.~(\ref{eq:hdN}).  The same 
will be true in our generalization to linearized gravity; we will rely on the higher-derivative form of the theory, analogous to Eq.~(\ref{eq:hdN}),
in the computations we present in Sec.~\ref{sec:examples} that illustrate the emergence of the nonlocal scale.  Nevertheless, we will present an auxiliary field 
formulation analogous to Eq.~(\ref{eq:start}), which is helpful in understanding the spectrum of massive Lee-Wick gravitons.

\section{Asymptotically nonlocal gravity}\label{sec:gravity}

Let us now construct an asymptotically nonlocal theory in the gravitational sector. We work in Cartesian coordinates and consider a small perturbation from $D$-dimensional 
Minkowski spacetime parametrized as
\begin{align}
g{}_{\mu\nu} = \eta{}_{\mu\nu} + 2 \, \kappa \, h{}_{\mu\nu} \, ,
\end{align}
where $\kappa = \sqrt{8\pi G}$, we set $\hbar = c = 1$, and we work in the particle physics metric signature $(+,-,\ldots,-)$.  We discuss the
generalization to the full, nonlinear theory in Sec.~\ref{sec:conc}.  As a warmup, recall that the leading-order Einstein-Hilbert Lagrangian can then be written 
compactly as
\begin{align}
\label{eq:lagr-eh}
\mathcal{L}_\text{EH} &= \frac{1}{2\kappa^2} \sqrt{-g} R = -\frac12 h{}_{\mu\nu} \mathcal{O}^{\mu\nu\rho\sigma} h{}_{\rho\sigma} + \mathcal{O}(\kappa) \, ,
\end{align}
where we defined the symbols\footnote{Here, and in what follows, all indices are raised and lowered with the Minkowski metric.}
\begin{align}
\begin{split}
\mathcal{O}^{\mu\nu}_{\rho\sigma} &\equiv \mathcal{O}^{\mu\nu\alpha\beta} \eta{}_{\rho\alpha} \eta{}_{\sigma\beta} = \left( \mathbb{1}^{\mu\nu}_{\rho\sigma} - \eta{}^{\mu\nu}\eta{}_{\rho\sigma} \right) \Box + \eta{}^{\mu\nu}\partial{}_\rho\partial{}_\sigma + \eta{}_{\rho\sigma}\partial{}^\mu\partial{}^\nu - \mathcal{C}^{\mu\nu}_{\rho\sigma} \, , \\
\mathbb{1}^{\mu\nu}_{\rho\sigma} &\equiv \frac12 \left(  \delta{}^\mu_\rho \delta{}^\nu_\sigma + \delta{}^\mu_\sigma \delta{}^\nu_\rho \right) \, , \label{eq:one} \\
\mathcal{C}{}^{\mu\nu}_{\rho\sigma} &\equiv \frac12 \left( \delta{}^\mu_\rho\partial^\nu\partial_\sigma + \delta{}^\mu_\sigma\partial^\nu\partial_\rho + \delta{}^\nu_\rho\partial^\mu\partial_\sigma + \delta{}^\nu_\sigma\partial^\mu\partial_\rho \right) \, ,
\end{split}
\end{align}
and $\Box \equiv \eta{}^{\mu\nu} \partial{}_\mu \partial{}_\nu$. Observe that the operator $\mathcal{O}^{\mu\nu\rho\sigma}$ satisfies
\begin{align}
\mathcal{O}^{\mu\nu\rho\sigma} &= \mathcal{O}^{\nu\mu\rho\sigma} = \mathcal{O}^{\mu\nu\sigma\rho} = \mathcal{O}^{\rho\sigma\mu\nu} \, , \\
\partial{}_\mu \mathcal{O}^{\mu\nu\rho\sigma} &= \partial{}_\nu \mathcal{O}^{\mu\nu\rho\sigma} = \partial{}_\rho \mathcal{O}^{\mu\nu\rho\sigma} = \partial{}_\sigma \mathcal{O}^{\mu\nu\rho\sigma} = 0 \, . \label{eq:transverse}
\end{align}
Under a gauge transformation associated with the infinitesimal diffeomorphism $2\kappa\xi^\mu(x)$, the metric perturbation transforms as $h{}_{\mu\nu} \rightarrow h{}_{\mu\nu} + \partial{}_\mu \xi{}_\nu + \partial{}_\nu \xi{}_\mu$. The gauge invariance of \eqref{eq:lagr-eh} is then guaranteed by Eq.~\eqref{eq:transverse}.

In order to construct an asymptotically nonlocal theory of linearized gravity, we proceed in close analogy to Eq.~(\ref{eq:start}): Consider a theory with $N$ 
fields $h{}_{\mu\nu}^j$ (where $j=1,\dots,N$) and $N-1$ auxiliary fields $\chi{}_{\mu\nu}^j$ (with $j=1,\dots,N-1$). Under a gauge transformation we demand 
$h{}^j_{\mu\nu} \rightarrow h{}^j_{\mu\nu} + \partial{}_\mu \xi{}_\nu + \partial{}_\nu \xi{}_\mu$ and $\chi{}^j_{\mu\nu} \rightarrow \chi{}^j_{\mu\nu}$. The Lagrangian is
\begin{align}
\label{eq:lagrangian}
\mathcal{L}_N = -\frac12 h{}^1_{\mu\nu} \mathcal{O}^{\mu\nu\rho\sigma} h{}^N_{\rho\sigma} - \sum\limits_{j=1}^{N-1} \chi_j^{\rho\sigma} \left[ \mathcal{O}^{\mu\nu}_{\rho\sigma} h{}^j_{\mu\nu} - m_j^2 \mathcal{M}{}^{\mu\nu}_{\rho\sigma}( h{}^{j+1}_{\mu\nu} - h{}^j_{\mu\nu} ) \right] \, ,
\end{align}
where the mass matrix $\mathcal{M}$ and its inverse $\mathcal{W}$ are
\begin{align}
\begin{split}
\mathcal{M}{}^{\mu\nu}_{\rho\sigma} &= \mathbb{1}^{\mu\nu}_{\rho\sigma} - b \, \eta{}^{\mu\nu}\eta{}_{\rho\sigma} \, , \\
\mathcal{W}{}^{\mu\nu}_{\rho\sigma} &= \mathbb{1}^{\mu\nu}_{\rho\sigma} - \frac{b}{bD-1} \eta{}^{\mu\nu}\eta{}_{\rho\sigma} \, , 
\end{split}
\label{eq:mandw}
\end{align}
such that $\mathcal{M}{}^{\mu\nu}_{\alpha\beta}\mathcal{W}{}^{\alpha\beta}_{\rho\sigma} = \mathbb{1}^{\mu\nu}_{\rho\sigma}$, with the parameter 
$b$ to be determined and left free for now, and $m_j^2>0$ are arbitrary mass parameters.

\subsection{Integrating out auxiliary fields}
The fields $\chi{}^j_{\mu\nu}$ appear linearly in the Lagrangian and are therefore auxiliary fields.  Hence, one may perform the functional integral over each one of these exactly, leading to the iterative functional constraints
\begin{align}
\mathcal{O}^{\rho\sigma}_{\mu\nu} h{}^j_{\rho\sigma} = m_j^2 \mathcal{M}{}^{\rho\sigma}_{\mu\nu}( h{}^{j+1}_{\rho\sigma} - h{}^j_{\rho\sigma} ) \, .
\end{align}
By acting with the inverse mass matrix $\mathcal{W}$ one finds
\begin{align}
h{}^{j+1}_{\mu\nu} = \left( \mathbb{1}^{\rho\sigma}_{\mu\nu} + \frac{1}{m_j^2} \hat{\mathcal{O}}{}^{\rho\sigma}_{\mu\nu} \right) h{}^j_{\rho\sigma} \, , \quad
\hat{\mathcal{O}}{}^{\rho\sigma}_{\mu\nu} \equiv \mathcal{W}{}^{\rho\sigma}_{\alpha\beta} \mathcal{O}^{\alpha\beta}_{\mu\nu} \, ,
\end{align}
where $\hat{\mathcal{O}}$ is given by
\begin{align}
\hat{\mathcal{O}}{}^{\mu\nu}_{\rho\sigma} =  \mathcal{O}^{\mu\nu}_{\rho\sigma} + \frac{b(D-2)}{bD-1} \eta{}^{\mu\nu} (\Box\eta{}_{\rho\sigma} - \partial{}_\rho\partial{}_\sigma) \, .
\end{align}
After integrating out all constraints one is left with the Lagrangian
\begin{align}
\mathcal{L}_N = -\frac12 h^1_{\mu\nu} \left[ \mathcal{O} \left( \mathbb{1} + \frac{1}{m_{N-1}^2} \hat{\mathcal{O}} \right) \left( \mathbb{1} 
+ \frac{1}{m_{N-2}^2} \hat{\mathcal{O}} \right) \dots \left( \mathbb{1} + \frac{1}{m_1^2} \hat{\mathcal{O}} \right) \right]^{\mu\nu\rho\sigma} h^1_{\rho\sigma} \, ,
\label{eq:intermediate}
\end{align}
where we have suppressed the tensor indices in the intermediate operators.  We find that the higher-derivative Lagrangian we seek is
obtained for the choice $b=1/2$.  This leads to significant simplifications, including the relations
\begin{align}
\label{eq:master-relations}
\hat{\mathcal{O}}{}^{\mu\nu}_{\alpha\beta} \hat{\mathcal{O}}{}^{\alpha\beta}_{\rho\sigma} = \Box \hat{\mathcal{O}}{}^{\mu\nu}_{\rho\sigma} \, , \qquad
\mathcal{O}{}^{\mu\nu}_{\alpha\beta} \hat{\mathcal{O}}{}^{\alpha\beta}_{\rho\sigma} = \Box \mathcal{O}{}^{\mu\nu}_{\rho\sigma} \, ,
\end{align}
which hold for any number of spacetime dimensions $D$.   These may be used to show that Eq.~(\ref{eq:intermediate}) simplifies to
\begin{align}
\label{eq:lagrangian-2}
\mathcal{L}_N = -\frac12 h^1_{\mu\nu} f(\Box) \mathcal{O}^{\mu\nu\rho\sigma} h^1_{\rho\sigma} \, , \quad f(\Box) \equiv \prod\limits_{j=1}^{N-1} \left( 1 + \frac{\Box}{m_j^2} \right) \, .
\end{align}
Compared to the linearized Einstein-Hilbert Lagrangian \eqref{eq:lagr-eh}, the above implements a higher-derivative modification thereof in terms of the form factor $f(\Box)$, where $h{}^1_{\mu\nu}$ is the graviton field.

It is worth noting that the auxiliary field formulation in  Eq.~(\ref{eq:lagrangian}) would not lead to the simple relations in 
Eq.~(\ref{eq:master-relations}), nor the desired end result in Eq.~(\ref{eq:lagrangian-2}), for a more general form 
of $\mathcal{M}{}^{\mu\nu}_{\rho\sigma}$.  This makes the present construction nontrivial and quite different from that of the scalar and gauge theory models considered in our prior work~\cite{Boos:2021chb,Boos:2021jih,Boos:2021lsj}.
The choice $b=1$ in Eq.~(\ref{eq:mandw}) corresponds to the tensor structure of  a Pauli-Fierz mass term; in models of
massive gravity~\cite{Hinterbichler:2011tt}, this is usually the preferred choice since it renders the massive graviton free of a ghost degree of 
freedom.  In the present context, we retain a massless graviton, and the additional massive states already include a proliferation of ghosts.  The
extra degree of freedom in each massive mode from the choice $b=1/2$ does nothing more than indicate that the Lee-Wick spectrum includes 
both spin-two and spin-zero Lee-Wick particles, with the latter quantized like any other Lee-Wick scalar~\cite{Park:2010zw}.   All are ultimately
decoupled as one takes that asymptotically nonlocal limit. 

\subsection{Propagator}
In order to find the graviton propagator we add a gauge-fixing term to the Lagrangian,\footnote{The required Faddeev-Popov ghosts do not appear in our subsequent computations.}
\begin{align}
\mathcal{L}_\text{gf} = \frac{1}{2\xi} \left( \partial{}_\mu h{}^{\mu\nu}_1 - \lambda \partial{}^\nu h_1 \right)^2 \, , \quad h_1 \equiv \eta{}^{\alpha\beta} h^1_{\alpha\beta} \, ,
\end{align}
which, after integration by parts, we may rewrite as
\begin{align}
\mathcal{L}_\text{gf} = -\frac{1}{2\xi} h{}^1_{\mu\nu} \left[ \lambda^2 \eta{}^{\mu\nu}\eta{}_{\rho\sigma} \Box - \lambda( \eta{}^{\mu\nu}\partial{}_\rho \partial{}_\sigma + \eta{}_{\rho\sigma}\partial{}^\mu\partial{}^\nu ) + \frac12 \mathcal{C}^{\mu\nu}_{\rho\sigma} \right] h{}^{\rho\sigma}_1 \, .
\end{align}
Then, for $\lambda=\tfrac12$, the propagator takes the form
\begin{align}
\GravitonPropagator &\equiv D{}^{\mu\nu\rho\sigma}(k) \nonumber \\
&= \frac{i}{2k^2 f(-k^2)} \bigg\{ \eta{}^{\mu\rho}\eta{}^{\nu\sigma}+\eta{}^{\mu\sigma}\eta{}^{\nu\rho} - \frac{2}{D-2}\eta{}^{\mu\nu}\eta{}^{\rho\sigma} \label{eq:gravprop} \\
&\hspace{40pt}-[1-2\xi f(-k^2)]\frac{\eta{}^{\mu\rho}k^\nu k^\sigma + \eta{}^{\mu\sigma}k^\nu k^\rho + \eta{}^{\nu\rho}k^\mu k^\sigma + \eta{}^{\nu\sigma}k^\mu k^\rho}{k^2} \bigg\} \, , 
\nonumber  \\
f(-k^2) &\equiv \prod\limits_{j=1}^{N-1} \left( 1 - \frac{k^2}{m_j^2} \right) \, .
\end{align}
In a local theory, where $f(-k^2)=1$, the result obtained by setting $\xi=1/2$ in Eq.~(\ref{eq:gravprop}) is usually said to be the 
graviton propagator in harmonic or de Donder gauge.  Note that  $i \, [k^2 f(-k^2)]^{-1}$ is identical to Eq.~(\ref{eq:nondegprop}) and
has the same partial fraction decomposition,
\begin{equation}
\frac{1}{k^2 \, f(-k^2)} = \sum_{k=0}^{N-1} \frac{c_j}{k^2 - m_j^2} \,\, ,
\end{equation}
with the coefficients $c_j$ given in Eq.~(\ref{eq:pfd0}).  For later convenience, it is useful to note that the $c_j$  satisfy the sum rules
\begin{align}
\label{eq:sum-rules}
\sum\limits_{j=0}^{N-1} c_j = 0 \,\,\,\, \mbox{ and } \,\,\,\,\, \sum\limits_{j=0}^{N-1} m_j^{2n} \, c_j = 0 ~ \quad \text{for} ~ n=1,\dots,N-2 \, .
\end{align}

\section{Examples of emergent scale}\label{sec:examples}

In this section, we present a number of examples that illustrate the appearance of the derived nonlocal scale in calculations involving 
the finite-$N$ theory.    We confirm that in the limit of large $N$, we recover known results from the literature on nonlocal 
gravity~\cite{Tomboulis:1997gg,Modesto:2011kw,Biswas:2011ar,Talaganis:2014ida,Modesto:2015ozb,Li:2015bqa,Modesto:2017sdr,Buoninfante:2018gce}.  
This is useful to verify that results obtained via the limiting procedure ({\it i.e.}, via a computation that assumes the theory \eqref{eq:hdN} with $N$ 
large but finite) and at the nonlocal limit point, Eq.~(\ref{eq:limitpoint}), do not exhibit any discontinuities.

\subsection{Metric of a classical point particle}
Taking the Lagrangian \eqref{eq:lagrangian-2}, we may substitute $h{}_{\mu\nu} \equiv h{}^1_{\mu\nu}$ for notational brevity and couple this to matter in the usual way,
\begin{align}
\mathcal{L} = \mathcal{L}_N - \kappa \, h{}_{\mu\nu} T{}^{\mu\nu} \, ,
\end{align}
such that the classical field equations take the form
\begin{align}
\mathcal{O}^{\rho\sigma}_{\mu\nu} f(\Box) h{}_{\rho\sigma} = - \kappa\, T{}_{\mu\nu} \, .
\end{align}
Working in the harmonic gauge, $\partial{}^\mu h{}_{\mu\nu} = \tfrac12 \partial{}_\nu h$, the field equations take the form
\begin{align}
\Box f(\Box) \left( h{}_{\mu\nu} - \frac12 h \eta{}_{\mu\nu}\right) = - \kappa \,T{}_{\mu\nu} \, ,
\end{align}
and energy-momentum conservation follows from the gauge choice directly. Let us find the weak-field solution for a point particle of mass $m$ at rest. Let $X{}^\mu = (t, \boldsymbol{x})$, the symbol $\boldsymbol{x}$ denote a $(D-1)$-dimensional spatial vector in Cartesian coordinates, and $r \equiv |\boldsymbol{x}|$. The conserved external energy-momentum tensor is
\begin{align}
T{}_{\mu\nu} = m \, \delta{}^t_\mu \delta{}^t_\nu \, \delta{}^{(D-1)}( \boldsymbol{x}) \, .
\end{align}
Inserting the static and spherically symmetric ansatz
\begin{align}
\dd s^2 = (\eta{}_{\mu\nu} + 2\kappa h{}_{\mu\nu})\dd X{}^\mu \dd X{}^\nu \equiv +[1-\phi(r)]\dd t^2 - [1 + \psi(r)]\dd\boldsymbol{x}^2
\end{align}
into the field equations yields
\begin{align}
\label{eq:twofield}
\begin{split}
f(\nabla^2) \nabla^2 \left[ \phi + (D-1)\psi \right] &= -4\kappa^2 m \, \delta^{(D-1)}(\boldsymbol{x}) \, ,  \\
f(\nabla^2) \nabla^2 \left[ (D-3)\psi - \phi \right] &= 0 \, ,
\end{split}
\end{align}
where $\nabla^2$ is the spatial part of the $\Box$-operator.  Equation~(\ref{eq:twofield}) is equivalent to
\begin{align}
f(\nabla^2) \nabla^2 \phi = -2 \, \frac{D-3}{D-2} \, \kappa^2 m\, \delta{}^{(D-1)}(\boldsymbol{x}) \, ,  
\label{eq:givesphi}
\end{align}
and $\psi = \phi/(D-3)$. One may verify that in the case where $D=4$ and $f=1$,  the solution of Eq.~(\ref{eq:givesphi}) is 
$\phi = 2\,G\,m/r$, as expected. In the more general case, we fix $D=4$ and use the partial fraction decomposition 
Eqs.~(\ref{eq:nondegprop})-(\ref{eq:pfd0}) to evaluate  the Fourier transform of Eq.~(\ref{eq:givesphi}).  
Solving for $\phi$ in this way, one finds
\begin{align}
\phi(r) = \psi(r) = \frac{2 G m}{r} \left[ 1 + \sum\limits_{j=1}^{N-1} c_j e^{-m_j r} \right] \, , \quad c_{j\ge 1} = -\prod\limits_{\substack{k=1\\k\not=j}}^{N-1} \frac{m_k^2}{m_k^2-m_j^2} \, ,
\label{eq:phiofr}
\end{align}
which resembles results encountered in quadratic gravity \cite{Stelle:1978,Quandt:1990gc,Modesto:2014eta}. Performing an expansion near $r=0$ (and
recalling that $c_0=1$ and $m_0=0$) one finds
\begin{align}
\phi(r) \approx 2Gm \left[ \frac{1}{r} \sum\limits_{j=0}^{N-1} c_j - \sum\limits_{j=0}^{N-1} c_j m_j + \frac{r}{2} \sum\limits_{j=0}^{N-1} c_j m_j^2 \right] + \mathcal{O}(r^2) \, .
\end{align}
The sum rules \eqref{eq:sum-rules} imply that for any $N \geq 2$ the $1/r$-divergence cancels, so the potential is manifestly finite at the 
origin. Moreover, for $N \ge 3$ the term linear in $r$ also vanishes, which implies the absence of a conical singularity at $r=0$; for a detailed study of regularity properties in higher-derivative gravity models see Ref.~\cite{Burzilla:2020utr}.

\begin{figure}[!ht]
\centering
\includegraphics[width=0.75\textwidth]{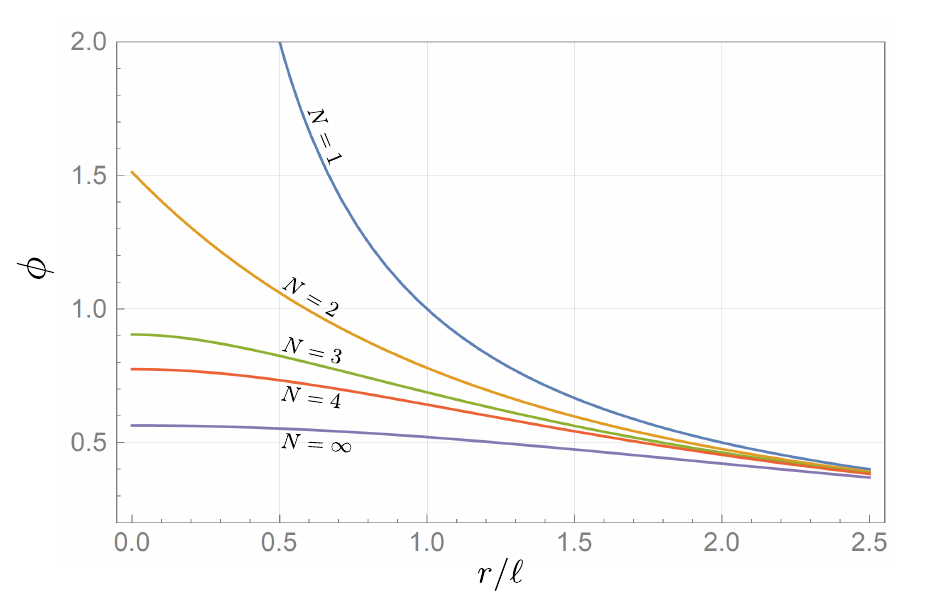}
\caption{The gravitational potential $\phi(r)$ for various choices of $N$, given a typical mass parametrization \eqref{eq:masses}, versus the dimensionless distance $r/\ell$, where $\ell$ is the emergent regulator scale. We choose $P=2$ and display the potentials for $2Gm/\ell = 1$. The case $N=1$ corresponds to the Newtonian potential, whereas $N \ge 2$ corresponds to the Lee-Wick case; $N=\infty$ is the nonlocal case.}
\label{fig:potentials} 
\end{figure}

The emergence of the nonlocal regulator scale can be seen by evaluating Eq.~(\ref{eq:phiofr}) numerically in the limit
where $N \rightarrow \infty$ with the $m^2_j/N$ approaching a common value $1/\ell^2$.   This is shown in Fig.~\ref{fig:potentials},
which assumes the following mass parametrization~\cite{Boos:2021lsj}:
\begin{align}
\label{eq:masses}
m_j^2 = \frac{N}{\ell^2} \frac{1}{1-\frac{j}{2N^P}} \, , \qquad j \geq 1 \,\,\, ,
\end{align}
where $P>1$ is an arbitrary parameter. The results can be seen to approach the expectation for a limiting nonlocal theory with 
an $\exp{(\ell^2 \Box)}$ form factor~\cite{Tseytlin:1995uq,Nicolini:2005vd,Modesto:2010uh,Edholm:2016hbt,Giacchini:2016xns,Boos:2018bxf,Akil:2022coa},
\begin{align}
\phi(r) = \psi(r) = \frac{2 G m}{r} \text{erf} \left( \frac{r}{2\ell} \right) \, ,
\end{align}
which is regular at $r=0$ and approaches the Newtonian expression for $r \gg \ell$.

\subsection{Nonrelativistic gravitational potential}
Using the propagator developed in Sec.~\ref{sec:gravity},  we next compute the gravitational potential by considering the nonrelativistic limit of a two-into-two
scattering amplitude.  To make the analogy with the well-known computation of the Coulomb potential in quantum electrodynamics manifest, we 
take the matter fields to be two distinct Dirac fermions with mass $m$.  The single-graviton vertex comes from the part of the Lagrangian that 
is linear in $h_{\mu\nu}$,
\begin{equation}
{\cal L} \supset - \kappa \, h_{\mu\nu} \, T^{\mu\nu} \, ,
\end{equation}
where
\begin{equation}
T^{\mu\nu} = \frac{i}{2} \overline{\psi} \stackrel{\leftrightarrow}{\partial}{}^{(\mu} \gamma^{\nu)} \psi-\eta^{\mu\nu} \left[\frac{i}{2} \, \overline{\psi} \stackrel{\!\!\! \leftrightarrow}{\partial_{\alpha}} \gamma^{\alpha} \psi - m \, \overline{\psi} \, \psi \right] 
\end{equation}
is the flat-space energy-momentum tensor.  Here we follow the conventions that $X_{(\mu\nu)} \equiv (X_{\mu\nu} + X_{\nu\mu})/2$ and 
$A \stackrel{\!\! \leftrightarrow}{\partial^{\mu}} B \equiv A \, \partial_\mu B - (\partial_\mu A) B$. The vertex Feynman rule is given by
\begin{equation}
\OneGravitonOneFermionVertex \equiv V(p',p)^{\mu\nu}= -i \, \kappa \left[ \frac{1}{2} \left( \mathbb{1}^{\mu\nu}_{\rho\sigma} - \eta_{\rho\sigma} \eta^{\mu\nu} \right) \gamma^\rho (p+p^\prime)^\sigma 
+ m \, \eta^{\mu\nu} \right] \, ,
\end{equation}
where $\mathbb{1}_{\rho\sigma}^{\mu\nu}$ is defined in Eq.~\eqref{eq:one}. To extract the gravitational potential, it is sufficient for us to study fermion-fermion scattering, whose scattering amplitude is given by
\begin{align}
\begin{split}
i {\cal M} &= \GravitonExchange = \left[ \overline{u}^{(s')}(p^\prime) V(p',p)_{\alpha\beta} \, u^{(s)}(p) \right] D^{\alpha\beta\rho\sigma}(q) \left[\overline{u}^{(r')}(k') V(k',k)_{\rho\sigma} \, u^{(r)}(k)\right]  \,\,\, ,
\label{eq:scatamp}
\end{split}
\end{align}
where $q=p-p'$ is the momentum flowing through the $t$-channel propagator, and $r$, $r'$, $s$, and $s'$ label spin states.  The portion of the propagator 
proportional to $[1-2 \xi f(-q^2)]$ gives a vanishing contribution to the amplitude, which can be verified using  $\overline{u}(p^\prime) \slash{\!\!\!q} \, u(p) = 0$ and $\overline{u}(k^\prime) \slash{\!\!\!q}  \, u(k)=0$.  The remaining part of the amplitude simplifies dramatically in the nonrelativistic limit.  At zeroth order in the three-momentum, the spinor $u(p)$ 
in the Weyl basis may be written as \cite{Peskin:1995ev} 
\begin{equation}
u^{(r)}(p)  = \sqrt{m} \left(\begin{array}{c} \xi^{(r)} \\ \xi^{(r)} \end{array}\right) \, ,
\end{equation}
where $\xi^{(r)}$ (with $r=1, 2$) is a set of two-component spinors that describe the spin state of the particle in the rest frame.   For example, at lowest order, 
\begin{equation}
\overline{u}^{(r)}(p^\prime) \gamma^0 \, u^{(s)}(p) = \overline{u}^{(r)} (p^\prime) \, u^{(s)}(p) = 2\, m \,\xi^{\prime (r) \dagger} \xi^{(s)} \, .
\end{equation}
The scattering amplitude in Eq.~(\ref{eq:scatamp}) reduces in the same limit to
\begin{equation}
i {\cal M} = - \frac{\kappa^2 \, m^2 }{2 \, f(|\vec{p}-\vec{p'}|^2)} \frac{i}{|\vec{p}-\vec{p'}|^2} \left(2\, m \,\xi^{\prime (s') \dagger} \xi^{(s)} \right) 
\left(2\, m \,\xi^{\prime (r') \dagger} \xi^{(r)} \right) \, .
\end{equation}
One may immediately identify the Fourier transformed potential energy~\cite{Peskin:1995ev},
\begin{equation}
\widetilde{V}(\vec{q}) = - 4 \, \pi \, G \, \frac{m^2}{|\vec{q}|^2 f(|\vec{q}|^2)} \, .
\end{equation}
We decompose $[\,|\vec{q}|^2 f(|\vec{q}|^2)\,]^{-1}$ using partial fractions and then Fourier transform,
\begin{equation}
V(\vec{x}) = -4\, \pi \, G \,m^2 \int \frac{\dd^3 q}{(2 \pi)^3} \, e^{i\vec{q}\cdot\vec{x}} \left[\frac{1}{|\vec{q}|^2}+\sum_{j=1}^{N-1} c_j \, \frac{1}{|\vec{q}|^2 + m_k^2} \right],
\end{equation}
where the $c_j$ are the same coefficients defined in Eq.~(\ref{eq:pfd0}).  Regulating the first term in the usual way, one obtains
\begin{equation}
V(\vec{x}) = - \frac{G \, m^2}{r}\left[1 + \sum_{j=1}^{N-1} c_j \, e^{-m_j r} \right] \, ,
\end{equation}
where $r \equiv | \vec{x}|$. This potential energy function is proportional to the function $\phi(r)$ discussed in the previous section.  Hence,
the singularity at the origin is eliminated and the potential energy is regulated by the same emergent scale in the asymptotically nonlocal
limit.

\subsection{Loop regulator}
As a final example, let us now demonstrate that the emergent scale also regulates the otherwise quadratically divergent self-energy of a real scalar field of mass $m$. The vertex Feynman rules are given by
\begin{align}
\OneGravitonOneScalarVertex \hspace{16pt} &=  -i \kappa \, \left[ p^\mu p'^\nu + p^\nu p'^\mu - (p\cdot p' - m^2)\,\eta{}^{\mu\nu} \right] \, , \\[10pt]
\TwoGravitonTwoScalarsVertex \hspace{5pt}  &=  4\, i \kappa^2 \, \Big[\mathbb{1}^{\mu\nu}_{\alpha\gamma} \mathbb{1}^{\rho\sigma}_{\beta\delta} \, \eta^{\gamma\delta} \, (p'^\alpha p^\beta + p'^\beta p^\alpha) 
\nonumber \\[-30pt]
& \hspace{40pt} -\frac{1}{2} \,(\mathbb{1}^{\mu\nu\rho\sigma} - \frac{1}{2}\, \eta^{\mu\nu}\eta^{\rho\sigma} )(p\cdot p' - m^2) \nonumber \\
& \hspace{40pt} -\frac{1}{2} \,(\mathbb{1}^{\mu\nu}_{\alpha\beta} \eta^{\rho\sigma} + \mathbb{1}^{\rho\sigma}_{\alpha\beta} \eta^{\mu\nu}) \, p'^\alpha p^\beta \Big] \, ,
\end{align}
where $\mathbb{1}^{\mu\nu\rho\sigma}$ is defined by Eq.~(\ref{eq:one}), with all indices raised using the Minkowski metric $\eta^{\alpha\beta}$.  The total self-energy at one loop is\footnote{The self-energy includes only one-particle irreducible diagrams. Note that our
expansion about flat spacetime assumes a vanishing cosmological constant and no gravitational tadpoles.}
\begin{align}
-i M^2(p^2) = \BubbleDiagram + \RainbowDiagram \equiv -i \left[ M^2_A(p^2) + M^2_B(p^2) \right] \, .
\label{eq:msqrd}
\end{align}
The physical scalar mass $p^2 = m_{\rm phys}^2$ is determined by the location of the propagator pole, {\it i.e.} it is the solution to $p^2-m^2 -M^2(p^2)=0$.   This 
makes $M^2(m_{\rm phys}^2)$ a quantity of interest; at the order we work in perturbation theory, this is equivalent to $M^2(m^2)$, which we now study.   For 
simplicity, we shall work in $\xi=0$ gauge.\footnote{The on-shell self-energy in Abelian and non-Abelian gauge theories is gauge invariant, with no dependence
on the parameter $\xi$.  In gravity, this is not the case, so that mass renormalization requires gauge-dependent counterterms.  Nevertheless, it can be shown 
that physical quantities, such as scattering amplitudes, remain independent of gauge~\cite{Antoniadis:1985ub}.  For an alternative approach using a
background field formalism which leads to gauge-invariant counterterms, see Ref.~\cite{Mackay:2009cf}.} On-shell, the second, ``rainbow" diagram gives
\begin{align}
-i M^2_B(p^2=m^2) = 2\kappa^2 m^2 \int \frac{\dd^4 k}{(2\pi)^4} \frac{m^2 k^2-4(k\cdot p)^2}{k^4(k^2-2p\cdot k)f(-k^2)} \, .
\end{align}
If one were to set $f=1$, this expression looks logarithmically divergent based on naive power counting, provided a factor of $k^2$ survives in the numerator 
of the integrand.  However, this is not quite the case:  After combining denominators and shifting the integration variable $ k \rightarrow k + {\rm shift}$, the 
leading term in the numerator may be replaced by $(1-4/D) \, k^2 m^2$, with vanishing coefficient in $D=4$.  Hence, this integral represents a finite correction, even 
before faster convergence is provided by $f(-k^2)$.   Therefore, we focus on the first diagram to see the appearance of the nonlocal scale as a regulator.  The 
first, ``bubble" diagram is given by
\begin{align}
-i M_A^2(p^2=m^2) = -6 \, \kappa^2 m^2 \int \frac{\dd^4 k}{(2\pi)^4} \frac{1}{k^2f(-k^2)} \, .
\end{align}
This expression is quadratically divergent for $f=1$, so let us now track the influence of the higher-derivative modification. Performing a partial fraction decomposition,
we may write
\begin{align}
\begin{split}
-i M_A^2(p^2=m^2) &= 6\, i\kappa^2 m^2 \sum\limits_{j=0}^{N-1} c_j \int \frac{\dd^4 k_E}{(2\pi)^4} \frac{1}{k_E^2+m_j^2} \\
&= 6i\kappa^2 m^2 \sum\limits_{j=1}^{N-1} c_j \left[ -\frac{m_j^2}{8\pi^2}\times\frac{1}{\epsilon} + \text{finite} \right] \, ,
\end{split}
\end{align}
where in the second line we have restored $m_0=0$ and evaluated the integral using dimensional regularization.  The $1/\epsilon$-divergences cancel due to the sum rules \eqref{eq:sum-rules}, such that
\begin{align}
-iM_A^2(p^2=m^2) &= \frac{6i\kappa^2}{(4\pi)^2} m^2 \sum\limits_{j=1}^{N-1} c_j m_j^2 \log m_j^2 \approx \frac{6i\kappa^2}{(4\pi)^2} m^2 M_\text{nl}^2 + \mathcal{O}\left(\frac{1}{N}\right) \, .
\end{align}
The last equality can be found in Eq.~(4.28) of Ref.~\cite{Boos:2021jih} and follows from the parametrization in Eq.~\eqref{eq:masses}, and holds numerically for $P \ge 1$. Therefore, the emergent scale $M_\text{nl}$ acts as the physical regulator for the gravitational corrections to the scalar self energy as the Lee-Wick
spectrum is appropriately decoupled.

\section{Conclusions}\label{sec:conc}

In Refs.~\cite{Boos:2021chb,Boos:2021jih,Boos:2021lsj}, we introduced a class of theories that interpolate between a Lee-Wick theory, with a finite
number of higher-derivative quadratic terms, and a ghost-free nonlocal theory, with infinite-derivative quadratic terms. We call this sequence of theories, with
ever increasing numbers of Lee-Wick particles, asymptotically nonlocal. As the number of Lee-Wick particles is increased, their spectrum is taken to decouple
in such a way that the Lagrangian approaches that of the nonlocal theory.   Since the nonlocal scale serves as a regulator in this limiting theory, one
can anticipate the emergence of a derived regulator scale in  the asymptotically nonlocal theories with large but finite $N$; this scale does not correspond to any
fundamental parameter in the Lagrangian of the finite-$N$ theory.  The derived regulator scale is hierarchically smaller that the lightest Lee-Wick partner, with the 
suppression in the squared cut off scale proportional to the number of propagator poles.  This provides motivation for studying these theories: more conventional 
Lee-Wick theories lose the ability to address the hierarchy problem as the lightest Lee-Wick particle is decoupled from the low-energy effective theory.  Asymptotically 
nonlocal theories allow this decoupling while still providing the desired cancellation of quadratic divergences. We know of no other Lee-Wick theories discussed in 
the literature that have this property.

As asymptotic nonlocality has been previously considered in scalar field theories~\cite{Boos:2021chb}, Abelian gauge theories~\cite{Boos:2021jih}, and 
non-Abelian gauge theories~\cite{Boos:2021lsj}, respectively, it is natural to ask how the same construction might be generalized to the gravitational sector,
so that all the fundamental interactions are treated in a similar way.  The present work 
takes the first step in this direction by developing an asymptotically nonlocal version of linearized gravity.   As in our earlier work~\cite{Boos:2021chb,Boos:2021jih,Boos:2021lsj}, we 
present a higher-derivative and an auxiliary-field formulation of the gravitational theory that preserve the diffeomorphism invariance of the linearized Einstein-Hilbert action. 
We then consider a number of examples that confirm the emergence of the nonlocal scale, namely, in resolving the singularity at the origin of the 
nonrelativistic gravitational potential and in regulating the divergences of graviton loop diagrams.   As with the quantum field theories that we previously 
studied~\cite{Boos:2021chb,Boos:2021jih,Boos:2021lsj}, we find that the emergent regulator scale is also suppressed relative to lightest Lee-Wick particle according to 
the relation $M_\text{nl}^2 \sim {\cal O}\left(m_1^2 / N \right)$.

Generalization of asymptotically nonlocal gravity to the fully nonlinear theory can be formulated most easily working with a higher-derivative formulation,
for example,
\begin{equation}
\mathcal{L} = \sqrt{-g} \left[\frac{1}{2\kappa^2} R + R F_1(\Box) R + R{}_{\mu\nu} F_2(\Box) R{}^{\mu\nu} + R{}_{\mu\nu\rho\sigma} F_3(\Box) R{}^{\mu\nu\rho\sigma} \right] \, , 
\label{eq:full}
\end{equation}
\begin{equation}
 F_k(\Box) = \prod\limits_{j=1}^{N-1} \left(1 - \frac{\ell_{k,j}^2 \Box}{N-1} \right) \, , \quad \Box = g{}^{\mu\nu} \nabla{}_\mu \nabla{}_\nu \, ,
\end{equation}
where the $\nabla{}_\mu$ are covariant derivatives.   Here, the $F_i(\Box)$ can be chosen so that Eq.~(\ref{eq:full}) reduces to 
Eq.~\eqref{eq:lagrangian-2} in the linearized approximation \cite{Modesto:2011kw,Biswas:2011ar}; see also Ref.~\cite{Frolov:2015usa}. Finding a compact expression for Eq.~(\ref{eq:full}) 
in terms on auxiliary fields is a more difficult task, but is not strictly necessary for studying any
relevant physics of interest in the full theory.

The present work on asymptotically nonlocal gravity concludes a series of papers that systematically develop the framework of 
asymptotic nonlocality.   In each of these investigations, we found an unusual relationship between the particle mass spectrum and the 
regulator scale, one that provides a new possibility for addressing the hierarchy problem. 
The natural extension to gravity considered here led us to identify an interesting class of Lee-Wick gravitational theories and provides motivation as well as a solid theoretical foundation for their further development and phenomenological study.

%%%%%%%%%%%%%%%%%%%%%%%%%%%%%%%%%%%%%%%%%%%%%%%%%%%%%%%%%%%
\begin{acknowledgments}  
We thank the NSF for support under Grants PHY-1819575 and PHY-2112460.
\end{acknowledgments}

% \appendix

\end{document}